\documentclass[english,pra,showpacs,twocolumn]{revtex4-1}
\usepackage[T1]{fontenc}
\usepackage[latin9]{inputenc}
\usepackage{xcolor}
\usepackage{amsmath}
\usepackage{amssymb}
\usepackage{graphicx}

\makeatletter

\newcommand{\lyxdot}{.}

\usepackage[colorlinks = true,
            linkcolor =purple,
            urlcolor  =magenta,
            citecolor = blue,
            anchorcolor = blue]{hyperref}

\makeatother

\usepackage{babel}
\begin{document}
\title{Quantum state engineering using weak measurement }
\author{Qiang Hu }
\author{Taximaiti Yusufu }
\email{taxmamat_84@sina.com}

\author{Yusuf Turek}
\email{yusufu1984@hotmail.com}

\affiliation{School of Physics and Electronic Engineering, Xinjiang Normal University,
Urumqi, Xinjiang 830054, China}
\date{\today}
\begin{abstract}
State preparation via postselected weak measurement in three wave
mixing process is studied. Assuming the signal input mode prepared
in a vacuum state, coherent state or squeezed vacuum state, separately,
while the idler input prepared in weak coherent state and passing
the medium characterized by the second-order nonlinear susceptibility.
It is shown that when the single photon is detected at one of the
output channels of idler beam's path, the signal output channel is
prepared in single-photon Fock state, single-photon-added coherent
state or single-photon-added squeezed vacuum state with very high
fidelity, depending upon the input signal states and related controllable
parameters. The properties including squeezing, signal amplification,
second order correlation and Wigner functions of the weak measurement
based output states are also investigated. Our scheme promising to
provide alternate new effective method for producing useful nonclassical
states in quantum information processing. 
\end{abstract}
\maketitle

\section{\label{sec:1}introduction}

New state generation and its optimization have significant importance
in quantum information processing \citep{Haroche2006,4,Chuang2010,RN1927,Zhang2014,RN1928}.
There have plenty of research works studied various quantum states
and proposed the schemes for generating them. The particular interests
has been devoted to Fock states \citep{PhysRevLett.70.762,PhysRevLett.88.143601,RN1929},
Schrodinger cat states \citep{PhysRevA.72.022320,75,RN1933,74,PhysRevLett.121.143602,RN1935,RN1936,RN1937,RN1934}
, squeezed states \citep{RN1932}, photon number states \citep{PhysRevLett.56.58,PhysRevA.36.4547,PhysRevA.39.3414,RN1930,Liu_2004,Waks2006,RN1931},
binomial states \citep{79,RN1940,78,76}, and squeezed state excitations
\citep{PhysRevLett.97.083604,PhysRevLett.101.233605,Liu2015,RN1938}.
Another interesting class of nonclassical states such as photon-added
coherent states \citep{PhysRevA.43.492}, photon-subtracted or -added
squeezed states \citep{PhysRevA.75.032104} that have been a subject
of interest since they also have potential applications in many related
quantum information processing \citep{C12,WANG20171393,RN1980,RN1981}.
Those states can be produced by repeated applications of photon creation
or annihilation operators \citep{PhysRevA.72.033822}, respectively,
on a given states \citep{142,PhysRevA.72.023820,PhysRevA.74.033813,PhysRevA.82.063833,xu2019}. 

We know that the purposing the feasible schemes to generate specific
quantum states and their implementations in the Lab are exciting and
challenging tasks to the researchers. In specific quantum state generation
processes we usually used the conditional measurement since it useful
to control the requested parameters to produce the desired quantum
states \citep{PhysRevA.34.3143,PhysRevA.38.3556,PhysRevLett.65.976,PhysRevLett.68.3424,PhysRevA.49.5078,PhysRevA.55.3184,PhysRevA.75.064302,PhysRevA.98.013809}.
The weak measurement proposed in 1988 \citep{PhysRevLett.60.1351}
by Aharonov, Albert, and Vaidman is a typical conditional measurement
characterized by postselection and weak value. The weak measurement
theory have various applications (see \citep{RevModPhys.86.307} and
references therein) and it recently widely used to the state optimization
problems \citep{KOFMAN201243,PhysRevA.85.012113,Turek2015}. One of
the author of this work studied the state optimization by using weak
measurement \citep{PhysRevA.92.022109,Turek2020,21} and showed that
the postselected weak measurement really can change the inherent properties
of the given states. Furthermore, in recent work \citep{PhysRevLett.107.133603},
they purposed a theoretical scheme to amplify the single-photon nonlinearity
using weak measurements implemented in cross-Kerr interaction medium
characterized by the third-order nonlinear susceptibility $\chi^{(3)}$
and its experimental realization is given in \citep{nature2015}.
On the other hand, Shikano and his collaborators \citep{Yu2017} studied
the generation of phase-squeezed optical pulses with large coherent
amplitudes by post-selection of single photon based on the same setup
of Ref.\citep{PhysRevLett.107.133603}. Those results also indicated
the potential usefulness of postselected weak measurement in quantum
state engineering processes. However, to our knowledge, the specific
quantum state generation via weak measurement has not been investigated
in detail in any literature, and it is worth to study. 

In this paper, we introduce a new scheme to generate some typical
nonclassical states such single-photon Fock states, single photon
added coherent (SPAC) state and single photon added squeezed vacuum
(SPASV) state in three optical wave mixing process via postselected
weak measurement \citep{PhysRevLett.60.1351}. In order to achieve
our goal, we consider the signal and idler beams as pointer (measuring
system) and measured system, respectively. We assume that initially
the measured system prepared in very weak coherent state while the
pointer (signal) state prepared in coherent or squeezed vacuum state.
The strong pump field is treated as classical and the weak coupling
between the pointer and measured system is realized by BBO nonlinear
crystal which can generate entanglement between them. By properly
choosing the pre- and post-selction states of measured system and
detecting one photon in one of the output of idler mode, the output
channel of the pointer is prepared in desired state with high purity
for controllable parameters. We found that if our input pointer state
is prepared in coherent (squeezed vacuum) state, then we can generate
SPAC (SPASV) state with very high fidelity accompanied by small successful
rate. Our results indicated that in our scheme we also can generate
single photon Fock state if the initial pointer state prepared in
vacuum state. To further confirm the identities of those generated
states we also investigate their related properties such as squeezing,
second order correlations and Wigner functions. Interestingly, we
found that the new generated SPAC state in our scheme have advantages
to increase the signal-to-noise ratio (SNR) in postselected weak measurement
over nonpostselected case. 

This paper is organized as follows. Section. \ref{sec:2}, presents
the basic scheme for generation of new nonclassical states in three
wave mixing process via postselected weak measurement technique. The
generation of SPAC and SPAVS states and their inherent properties
are discussed in Section. \ref{sec:3} and Section.\ref{sec:4}, respectively.
In Section. \ref{sec:3}, we also investigate the advantages of postselected
weak measurement in signal amplification process over nonpostselected
case for SPAC state by adjusting the weak value of measured system
observables. Finally, a summary and concluding remarks are given in
Section. \ref{sec:5}.

\section{\label{sec:2}Model setup for the new state generation via postselected
weak measurement }

The Hamiltonian of a three-wave mixing device \citep{Int}, under
the rotating wave approximation (RWA), neglecting external drive and
signal fields, is 
\begin{equation}
H=\hbar\omega_{s}a^{\dagger}a+\hbar\omega_{i}b^{\dagger}b+\hbar\omega_{p}c^{\dagger}c+i\hbar\chi^{(2)}\left(a^{\dagger}b^{\dagger}c-abc^{\dagger}\right),\label{eq:1}
\end{equation}
where $a,$ $b$ and $c$ are the annihilation operators of the signal,
idler and pump with frequencies $\text{\ensuremath{\omega_{s}} }$,
$\omega_{i}$ and $\omega_{p}$, and $\chi^{(2)}$ is the coupling
strength characterized by a second-order nonlinear susceptibility
of BBO.This Hamiltonian can describe the process of nondegenerate
parametric down-conversion whereby a photon of the pump field is converted
into two photons, one for each of the modes $a$ and $b$ \citep{Int}.
Using the parametric approximation, assuming that the pump field to
be a strong coherent state of the form $\vert\gamma e^{-i\omega_{p}t}\rangle$,
then we can rewrite the above Hamiltonian in interaction picture with
$\omega_{p}=\omega_{i}+\omega_{s}$ as 
\begin{align}
H_{I} & =i\hbar\eta\left(a^{\dagger}b^{\dagger}-ab\right),\label{eq:2}
\end{align}
where $\eta=\gamma\chi^{(2)}$. Further, the above Hamiltonian is
equivalent to 
\begin{align}
H_{I} & =\hbar g\left(A\otimes p-B\otimes q\right),\label{eq:3}
\end{align}
if we introduce 
\begin{equation}
B=\frac{i}{\sqrt{2}}\left(b-b^{\dagger}\right),\ \ \ A=\frac{1}{\sqrt{2}}\left(b^{\dagger}+b\right),\label{eq:4}
\end{equation}
 and 
\begin{equation}
q=\frac{1}{\sqrt{2}}\left(a^{\dagger}+a\right),\ \ \ p=\frac{i}{\sqrt{2}}\left(a^{\dagger}-a\right),\label{eq:5}
\end{equation}
with $[A,B]=i$ and $[q,p]=i$, respectively. The two terms in Hamiltonian,
Eq. ( \ref{eq:3}), are in the forms we usually used in weak measurement
problems \citep{PhysRevLett.60.1351}. In this work we take the signal
beam with variables $q$ and $p$ is pointer, and idler beam with
variables $A$ and $B$ is measured system, respectively. 

\begin{figure}
\includegraphics[width=8cm]{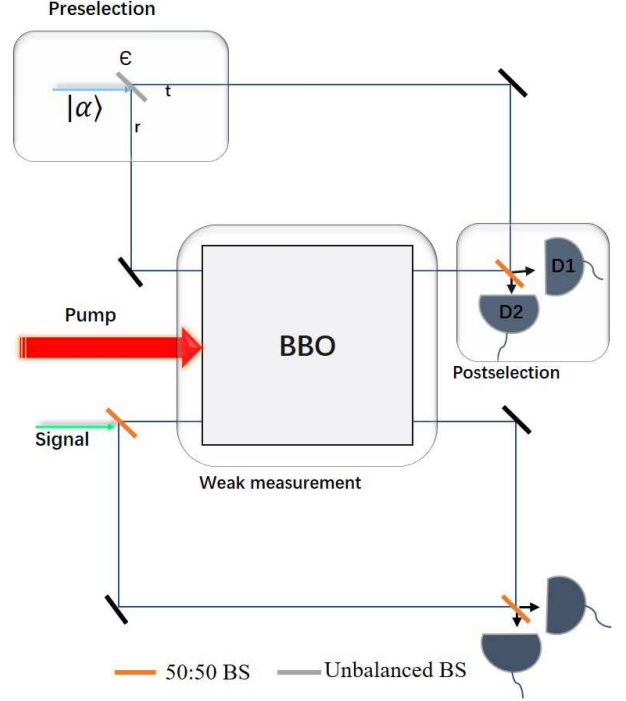}

\caption{\label{Fig:1}(Color online) Schematics of non-Gaussian state generation
in three wave mixing process using postselected weak measurement.
This model consists two Mach--Zehnder interferometers. The signal
and idler beams acts as pointer and measured system, respectively.
The preselection state prepared by weak coherent state $\vert\alpha\rangle$
passing through the unbalanced beam splitter (BS) with deviation $\epsilon$,
and signal beam initially prepared in some specific sates. The BBO
crystal playing the role for realizing the weak interaction between
pointer and measured system.The 50:50 BS in the upper Mach--Zehnder
interferometers takes the role of postselction, and the desired conditional
quantum state is generated in the output mode of signal beam after
we detect one photon by second photon detector (D2) in idler beam's
path. }
\end{figure}

The schematic setup of our state generation model is showed in Fig.\ref{Fig:1}.
As we can see from Fig.\ref{Fig:1}, there have two Mach--Zehnder
interferometers in our setup and the beam splitters are taking very
important roles to the implementation of our scheme. Beam splitters
are basic manipulations in classical and quantum optics to splitting
and mixing the optical beams. The input and output relations of beam
splitters can be described by Lie algebra \citep{PhysRevA.33.4033}.
In the Heisenberg picture, the photon annihilation operators of output
beam, $b_{k}$($k=1,2$) can be connected to input beam's annihilation
operators, $a_{k}$, as 
\begin{equation}
b_{k}=\sum_{j=1}^{2}U_{kj}a_{j},
\end{equation}
where $U_{kj}$ is the element of the scattering matrix 
\begin{equation}
U=\left(\begin{array}{cc}
\cos\vartheta e^{i\varphi_{t}}\  & \sin\vartheta e^{i\varphi_{r}}\\
-\sin\vartheta e^{-i\varphi_{r}} & \cos\vartheta e^{-i\varphi_{t}}
\end{array}\right).
\end{equation}
Here, $T=\cos\vartheta e^{i\varphi_{t}}$and $R=\sin\vartheta e^{i\varphi_{r}}$
are transmittance and reflectance of the beam splitter, respectively.
If $\varphi_{r}=\varphi_{t}=0$ and $\vartheta=\frac{\pi}{4}$, then
it becomes $50:50$ beam splitter. 

We assume that initially the measured system (idler beam) prepared
in weak coherent state with small amplitude ($\alpha\ll1$), and signal
beam prepared in some specific states such as squeezed vacuum state
and coherent state separately. In the upper optical path of our scheme,
we assume that the first beam splitter is slightly imbalanced with
small deviate $\epsilon$ to $50:50$ so that the preselection state
of the measured system can be written as \citep{PhysRevLett.108.080403}
\begin{equation}
\vert\psi_{i}\rangle=\vert\frac{\alpha}{\sqrt{2}}(1-\epsilon)\rangle_{t}\vert\frac{i\alpha}{\sqrt{2}}(1+\epsilon)\rangle_{r}.\label{eq:6}
\end{equation}
Here, the subscripts $t$ and $r$ indicates the transmitted and reflected
beams from the beam splitter. Then, the three wave mixing is realized
by the nonlinear BBO crystal which play the role to implement the
weak measurement process. In this process, the input photon annihilates
and produces two new mutually entangled photons. The unitary evolution
operator corresponding to the interaction Hamiltonian $H_{I}=\hbar g\left(A\otimes p-B\otimes q\right)$
which can implemented by BBO crystal is 
\begin{equation}
U=\exp\left(-\frac{i}{\hbar}\int_{0}^{t}H_{I}d\tau\right)=\exp\left(-ig\left[A\otimes p-B\otimes q\right]\right),\label{eq:U}
\end{equation}
where $g=\eta t$. Actually, this is the squeezing operator can generate
the two-mode vacuum squeezed state \citep{Int}. Here, $g$ is can
be considered as squeezing parameter which depends on pump intensity,
the crystal length, and its nonlinear coefficients. Following the
experimental work \citep{PhysRevA.82.063833}, we set $g=0.105$ throughout
this work. We can then rewrite the above unitary evolution operator
$U$ as 
\begin{equation}
U\simeq I-ig\left(A\otimes p-B\otimes q\right).\label{eq:8-1}
\end{equation}
 If we assume that the initial state of system and pointer are $\vert\psi_{i}\rangle$
and $\vert\phi\rangle$, after the unitary evolution the total system
state becomes as 
\begin{align}
\vert\Psi\rangle & =U\vert\psi_{i}\rangle\otimes\vert\phi\rangle\approx\left[I-ig\left(A\otimes p-B\otimes q\right)\right]\vert\psi_{i}\rangle\otimes\vert\phi\rangle.\label{eq:9-1}
\end{align}
 This is total system state before arrive to the second beam splitters
in our model (see Fig. \ref{Fig:1}). In our scheme the second splitters
are 50:50 with 50\% transmission and 50\% reflection. We take a postselection
to the idler beam accomplished by detectors in the upper optical paths.
Assume that the second photon detector (D2) is detect one photon and
the first photon detector (D1) no click i.e., $\vert1\rangle_{2d}\vert0\rangle_{1d}$.
This postselection process can be described by 

\begin{align}
|\psi_{f}\rangle_{2d} & =a_{2d}^{\dagger}|0\rangle_{r}|0\rangle_{t}\nonumber \\
 & =\text{\ensuremath{\frac{1}{\sqrt{2}}(|0\rangle_{r}|1\rangle_{t}-i|1\rangle_{r}|0\rangle_{t})},}\label{eq:pi}
\end{align}
where \begin{subequations}
\begin{align}
a_{2d} & =\frac{1}{\sqrt{2}}\left(a_{t}+ia_{r}\right),\label{eq:11}\\
a_{1d} & =\frac{1}{\sqrt{2}}\left(ia_{t}+a_{r}\right),
\end{align}
 \end{subequations}are the field operators relations between input
and output modes of the beam-splitter stransformation. After taking
the postselection with the postselected state $\vert\psi_{f}\rangle$
onto Eq. (\ref{eq:9-1} ), we can obtain the non normalized form of
the final state of the pointer (signal beam) and it reads as 
\begin{align}
\vert\Phi\rangle & =\langle\psi_{f}\vert\psi_{i}\rangle\left[1-ig\left(\langle A\rangle_{w}p-\langle B\rangle_{w}q\right)\right]\vert\phi\rangle\nonumber \\
 & =\langle\psi_{f}\vert\psi_{i}\rangle\left[1-\frac{g\alpha}{\sqrt{2}}a-\frac{g}{\sqrt{2}\alpha\epsilon}a^{\dagger}\right]\vert\phi\rangle,\label{eq:9}
\end{align}
where

\begin{align}
\langle A\rangle_{w} & =\frac{\langle\psi_{f}\vert\hat{A}\vert\psi_{i}\rangle}{\langle\psi_{f}\vert\psi_{i}\rangle}=\frac{\alpha}{2}-\frac{1}{2\alpha\epsilon},\label{eq:a}
\end{align}
and 
\begin{align}
\langle B\rangle_{w} & =\frac{\langle\psi_{f}\vert\hat{B}\vert\psi_{i}\rangle}{\langle\psi_{f}\vert\psi_{i}\rangle}=\frac{i}{2\alpha\epsilon}+i\frac{\alpha}{2},\label{eq:b}
\end{align}
are the weak values of $A$ and $B$, respectively. The probability
of finding one photon at D2 and no photon at D1 is 
\begin{equation}
P_{s}=\vert\langle\psi_{f}\vert\psi_{i}\rangle\vert^{2}=\vert\alpha\epsilon\vert^{2}.\label{eq:p}
\end{equation}
As we can see, the success probability of postselection $P_{s}$ is
depends on the imbalance $\epsilon$ caused by the little difference
between the reflection and transmission coefficients of the beam splitter
in the upper interferometer and weak coherent state amplitude $\alpha$
of the idler input state. From the Eqs. (\ref{eq:a}) and (\ref{eq:b}),
it can be seen that the weak values are generally complex, and can
take large values when the pre-selected state $\vert\psi_{i}\rangle$
and post-selected states$\vert\psi_{f}\rangle$ are almost orthogonal.
The magnitudes of weak idler input state amplitude $\alpha$, beam
splitter's deviation $\epsilon$, and coupling coefficient $g$ are
all controllable in optical experiments. Thus, we can manipulate and
change the inherent properties of the output signal state $\vert\Phi\rangle$
by adjusting these parameters. In the remaining parts of the work,
we study the new state generation and its verification processes by
taking the initial signal input sate $\vert\phi\rangle$ as coherent
state and vacuum squeezed state, respectively. 

\section{\label{sec:3}Generation of SPAC state}

In this section, we assume that the initial signal input state is
prepared as coherent state which defined as,
\begin{equation}
\vert\phi\rangle=\vert\beta\rangle=D(\beta)\vert0\rangle,\label{eq:co}
\end{equation}
where $\beta=\vert\beta\vert e^{i\theta}$ is complex number. For
this case, the output state of the signal, i.e., Eq. ( \ref{eq:9}),
is reads as 
\begin{equation}
\vert\varTheta\rangle=\mathcal{N}\left[\kappa_{1}\vert\beta\rangle-\kappa_{2}a^{\dagger}\vert\beta\rangle\right].\label{eq:post}
\end{equation}
Here, 
\begin{equation}
\mathcal{N}=\left[\vert\kappa_{1}\vert^{2}+\vert\kappa_{2}\vert^{2}(1+\vert\beta\vert^{2})-2Re[\kappa_{1}\kappa_{2}^{\ast}\beta]\right]^{-\frac{1}{2}},\label{eq:10}
\end{equation}
is the normalization constant, $\kappa_{1}=1-\frac{g\beta\alpha}{\sqrt{2}}$
and $\kappa_{2}=\frac{g}{\sqrt{2}\alpha\varepsilon}$, respectively.
It is very clear from Eq. (\ref{eq:post} ) that the output signal
state is a superposition of coherent state $\vert\beta\rangle$ and
SPAC state $a^{\dagger}\vert\beta\rangle$. As aforementioned, since
the all parameters $g$, $\alpha$ , $\epsilon$ and $\beta$ are
adjustable, the dominance of coherent state $\vert\beta\rangle$ and
SPAC state $a^{\dagger}\vert\beta\rangle$ can be completely controlled
by adjusting the related parameters. From the Eq. ( \ref{eq:post}),
we can see that if $\kappa_{2}\gg\kappa_{1}$ the state $\vert\varTheta\rangle$
reduced to the SPAC state $\vert1,\beta\rangle=\frac{a^{\dagger}\vert\beta\rangle}{\sqrt{1+\vert\beta\vert^{2}}}$.
In next sub sections we extend the discussions about properties of
the conditional output state $\vert\varTheta\rangle$.

\subsection{State Distance}

In quantum information theory, the quantification of the distance
of two quantum states described by density operators $\rho$ and $\sigma$
can be characterized by the quantum fidelity (or the called Uhlmann-Jozsa
fidelity) which is defined as 
\begin{equation}
F=\left(Tr\sqrt{\sqrt{\rho}\sigma\sqrt{\rho}}\right)^{2}.\label{eq:F}
\end{equation}
If both states are pure i.e., $\rho=\vert\psi\rangle\langle\psi\vert$
and $\sigma=\vert\phi\rangle\langle\phi\vert$, then 
\begin{equation}
F=\vert\langle\psi\vert\phi\rangle\vert^{2}.\label{eq:19}
\end{equation}
 This quantity is indeed a natural candidate for the state distance
since it corresponds to the closeness of states in the natural geometry
of Hilbert space. If $F=0$, the states are orthogonal or called totally
different (i.e., perfectly distinguishable). If $F=1$, then the two
states are totally same, $\vert\psi\rangle=\vert\phi\rangle$. 

\begin{figure}
\includegraphics[width=8cm]{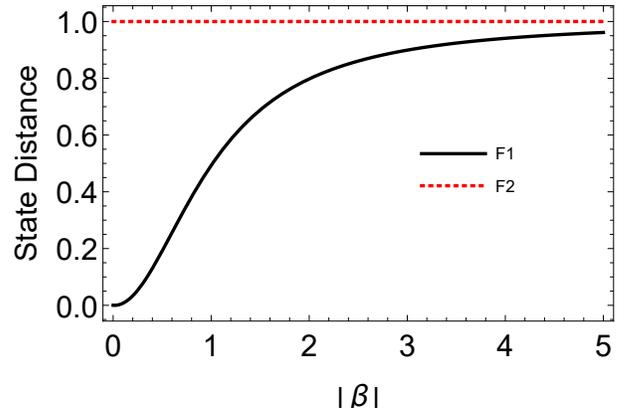}

\caption{\label{fig:2}(Color online)\textcolor{orange}{{} }State distance between
coherent state $\vert\beta\rangle$, SPAC state $\frac{a^{\dagger}\vert\beta\rangle}{\sqrt{1+\vert\beta\vert^{2}}}$
and $\vert\Phi\rangle$. Here we take $\theta=0$, $g=0.105$, $\alpha=0.01$
and $\epsilon=0.1$.}
\end{figure}

Here, in order to study the similarity of the output signal state
$\vert\varTheta\rangle$ between coherent state $\vert\beta\rangle$
and normalized SPAC state $\vert1,\alpha\rangle$, the fidelity between
$\vert\beta\rangle$, $\vert1,\alpha\rangle$ and $\vert\varTheta\rangle$
are calculated, and the result are given by 

\begin{equation}
F_{1}=\vert\langle\beta\vert\varTheta\rangle\vert^{2}=\vert\mathcal{N}(\kappa_{1}-\kappa_{2}\beta^{\ast})\vert^{2},
\end{equation}
 and 
\begin{align}
F_{2} & =\vert\langle1,\alpha\vert\varTheta\rangle\vert^{2}=\frac{\vert\mathcal{N}\vert^{2}\vert\kappa_{1}\beta-\kappa_{2}(1+\vert\beta\vert^{2})\vert^{2}}{1+\vert\beta\vert^{2}},
\end{align}
respectively. In Fig. \ref{fig:2}, we plot the fidelity $F_{1}$
and $F_{2}$ as a function of coherent state parameter $\vert\beta\vert$
for other fixed system parameters. As showed in Fig. \ref{fig:2},
the red dashed line shows the closeness between the output signal
state and the SPAC state, and it can be seen that the fidelity of
these two states always keeping the constant value ($F=1$) for all
$\vert\beta\vert.$ The Fig. \ref{fig:2} also indicated that the
$F_{1}$ is increased from zero to unity as $\vert\beta\vert$ increasing.
It can be seen that when $\alpha,\epsilon$ are much less than one
and $\vert\beta\vert$ is smaller, we can deduce that $\kappa_{2}\gg\kappa_{1}$.
Under this condition our generated output signal state is exactly
the SPAC state. 

\subsection{Second order correlation and Mandel factor}

Here, we study the second-order correlation function $g^{(2)}(0)$
and Mandel factor $Q_{m}$ of our generated signal state $\vert\varTheta\rangle$.
The second order correlation function of a single-mode radiation field
is defined as 
\begin{equation}
g^{(2)}\left(0\right)=\frac{\langle a^{\dagger2}a^{2}\rangle}{\;\langle a^{\dagger}a\rangle^{2}}.
\end{equation}
Its relations with the Mandel factor $Q_{m}$is 
\begin{equation}
Q_{m}=\langle a^{\dagger}a\rangle\left[g^{(2)}\left(0\right)-1\right].
\end{equation}
 If $0\le g^{(2)}\left(0\right)<1$ and $-1\le Q_{m}<0$ simultaneously,
the corresponding radiation field has sub-Poissonian statistics and
more nonclassical. We have remember that the Mandel factor $Q_{m}$
can never be smaller than $-1$ for any radiation fields, and negative
$Q_{m}$ values, which are equivalent to sub-poissonian statistics,
cannot be produced by any classical field.

The second-order correlation function $g^{(2)}(0)$ and Mandel factor
$Q_{m}$ of our generated output signal state $\vert\varTheta\rangle$
are given as \citep{Agarwal2013} 

\begin{equation}
g^{(2)}(0)=\frac{\langle\varTheta|a^{\dagger}a^{\dagger}aa|\varTheta\rangle}{\langle\varTheta|a^{\dagger}a|\varTheta\rangle^{2}},
\end{equation}
and 
\begin{align}
Q_{m} & =\langle\varTheta|a^{\dagger}a|\varTheta\rangle[g^{2}(0)-1],
\end{align}
respectively, with 

\begin{align}
\langle\varTheta|a^{\dagger2}a^{2}|\varTheta\rangle & =\mathcal{\vert N}\vert^{2}\{\vert\kappa_{1}\vert^{2}\vert\beta\vert^{4}-2\kappa_{2}^{\ast}\kappa_{1}Re[(2\vert\beta\vert^{2}\beta+\vert\beta\vert^{4}\beta]\nonumber \\
 & +\vert\kappa_{2}\vert^{2}(5\vert\beta\vert^{4}+\vert\beta\vert^{6}+4\vert\beta\vert^{2})\},\label{eq:aa}
\end{align}
and 

\begin{align}
\langle\varTheta|a^{\dagger}a|\varTheta\rangle & =\mathcal{\vert N}\vert^{2}\{\vert\kappa_{1}\vert^{2}\vert\beta\vert^{2}-2\kappa_{2}^{\ast}\kappa_{1}Re(\beta+\vert\beta\vert^{2}\beta)\nonumber \\
 & +\vert\kappa_{2}\vert^{2}(3\vert\beta\vert^{2}+\vert\beta\vert^{4}+1)\}.\label{eq:a-1}
\end{align}

\begin{figure}
\includegraphics[width=8cm]{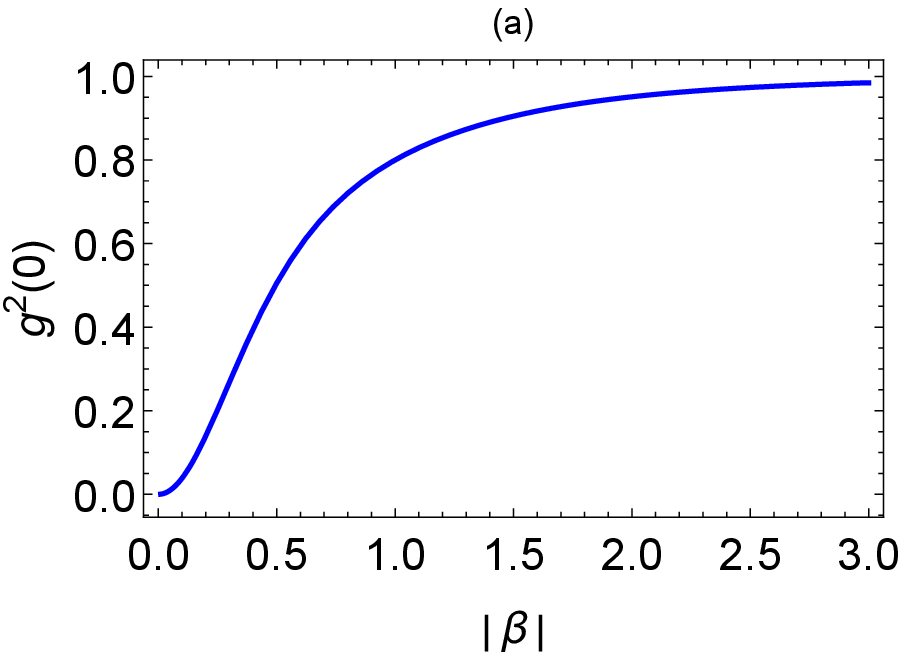}

\includegraphics[width=8cm]{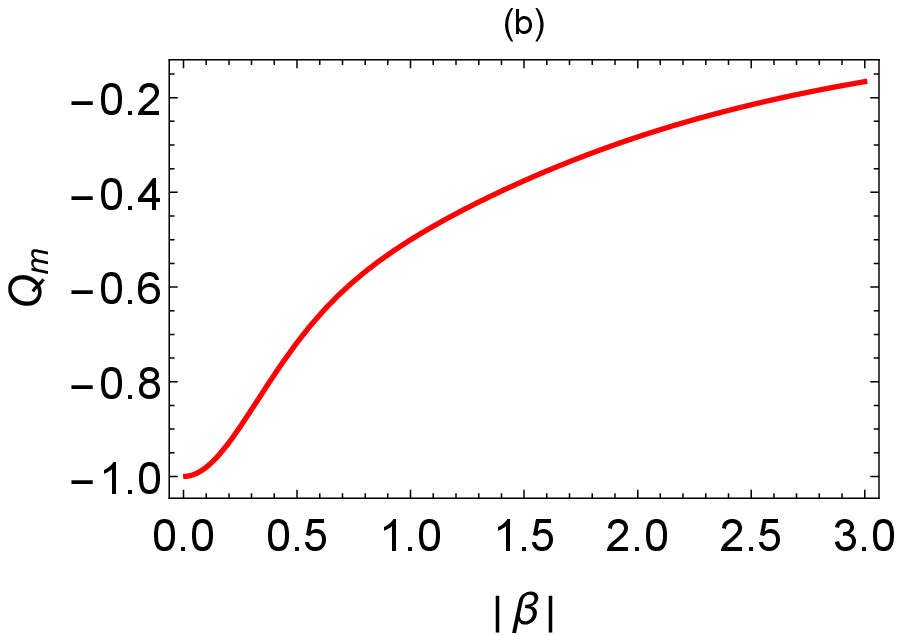}

\caption{\label{fig:4}(Color online) Second order correlation $g^{(2)}(0)$
and Mandel factor $Q_{m}$ of our generated signal output state $\vert\varTheta\rangle$
as a function of coherent state parameter $\vert\beta\vert$. (a)
$g^{(2)}\left(0\right)$ is varied. (b) $Q_{m}$ is varied. Other
parameters are the same as Fig. \ref{fig:2}}
\end{figure}

In Fig. \ref{fig:4}, we plot $g^{(2)}$$\left(0\right)$ and $Q_{m}$
as functions of coherent state parameter $\beta$ by fixing other
paramters to $\theta=0,g=0.105,\alpha=0.01$ and $\epsilon=0.1$.
As observed in Fig. \ref{fig:4}, $0\leq g^{(2)}(0)<1$ and$-1\leq Q_{m}<0$
for all plotted regions.This means that our generated signal output
field have sub-Poisson statistics which only possessed in nonclassical
states. Actually, the curves showed in Fig. \ref{fig:4} are matched
well with the corresponding curves of SPAC state $\vert1,\alpha\rangle$
\citep{PhysRevA.43.492}. Thus, we can further verified that in our
scheme we could effectively generate the SPAC state if the initial
signal input state is in coherent state with moderate parameter $\beta$. 

\subsection{Winger function }

To further verify our claim, in this subsection, we investigate the
Wigner function of $\vert\varTheta\rangle$. A state of a quantum
mechanical system is completely described by density matrix of a phase
space distribution such as the Wigner function. Every state function
has it unique phase space distributions and the Wigner distribution
function is the closest quantum analogue of the classical distribution
function in phase space. By evaluating the Wigner function we can
intuitively determine the strength of corresponding quantum nature,
and most importantly the negative value of the Wigner function can
prove the non-classicality of the state. In general, the Wigner function
is defined as the two-dimensional Fourier transform of the symmetric
order characteristic function, and the Wigner function for the state
$\rho=\vert\varTheta\rangle\langle\varTheta\vert$ can be written
as \citep{Int} 
\begin{equation}
W(z)\equiv\frac{1}{\pi^{2}}\int_{-\infty}^{+\infty}\exp(\lambda^{\ast}z-\lambda z^{\ast})C_{N}(\lambda)e^{-\frac{\lambda^{2}}{2}}d^{2}\lambda,\label{eq:35-1}
\end{equation}
where $C_{N}(\lambda)$ is the normal ordered characteristic function,
and is defined as 
\begin{equation}
C_{N}(\lambda)=Tr\left[\rho e^{\lambda a^{\dagger}}e^{-\lambda a}\right].\label{eq:34}
\end{equation}
After some calculation we can get the explicit expression of the Wigner
function of the state $\vert\varTheta\rangle$ and it given as 

\begin{align}
W(z) & =\!\frac{2\vert\mathcal{N}\vert^{2}}{\pi}\!\{\vert\text{\ensuremath{\kappa}}_{1}\vert^{2}\!e^{-2\vert z-\beta\vert^{2}}\!-\text{\!}\vert\kappa_{2}\vert^{2}\!\left(1\!-\text{\!\!}\vert2z-\beta\vert^{2}\right)e^{-2\vert z-\beta\vert^{2}}\nonumber \\
 & -Re\left[\kappa_{2}\kappa_{1}^{\ast}\left(2Re[\beta]-z\right)e^{\frac{1}{2}\left[(z-\beta)^{2}+(z^{\ast}-\beta^{\ast})^{2}\right]}\right]\}.\label{eq:18}
\end{align}

\begin{figure}
\includegraphics[width=8cm]{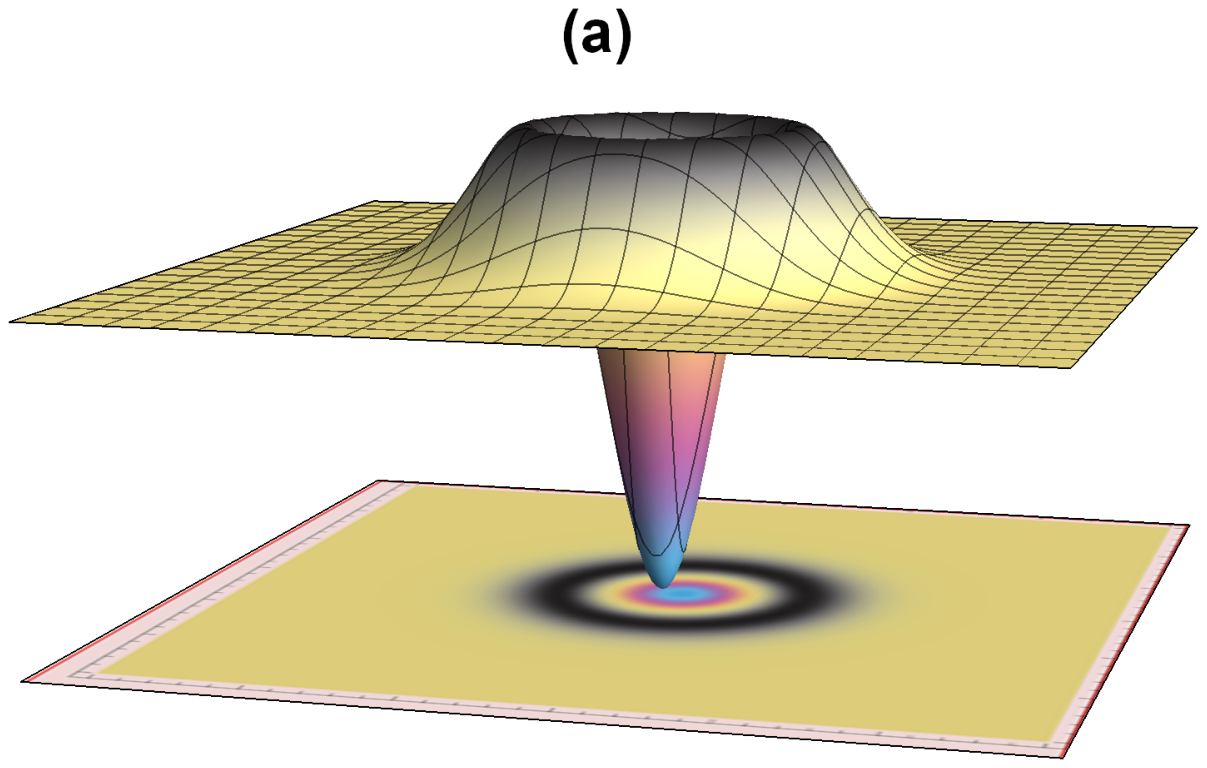}

\includegraphics[width=8cm]{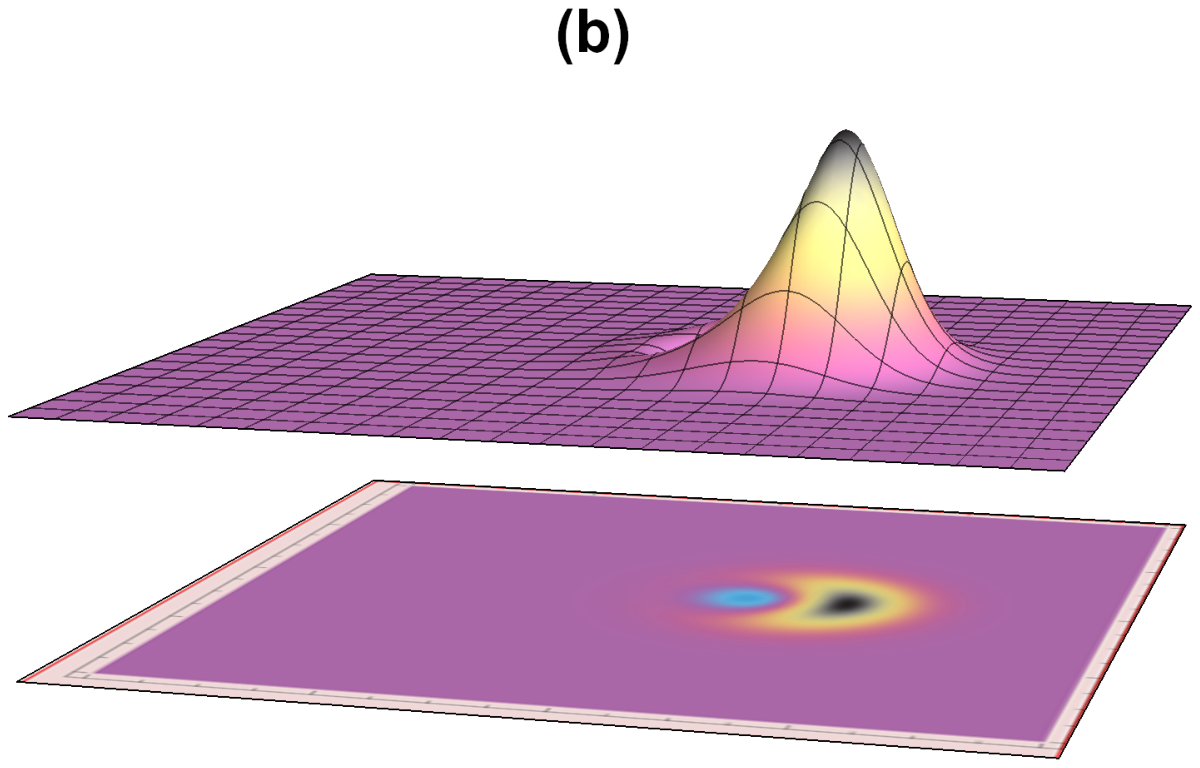}

\includegraphics[width=8cm]{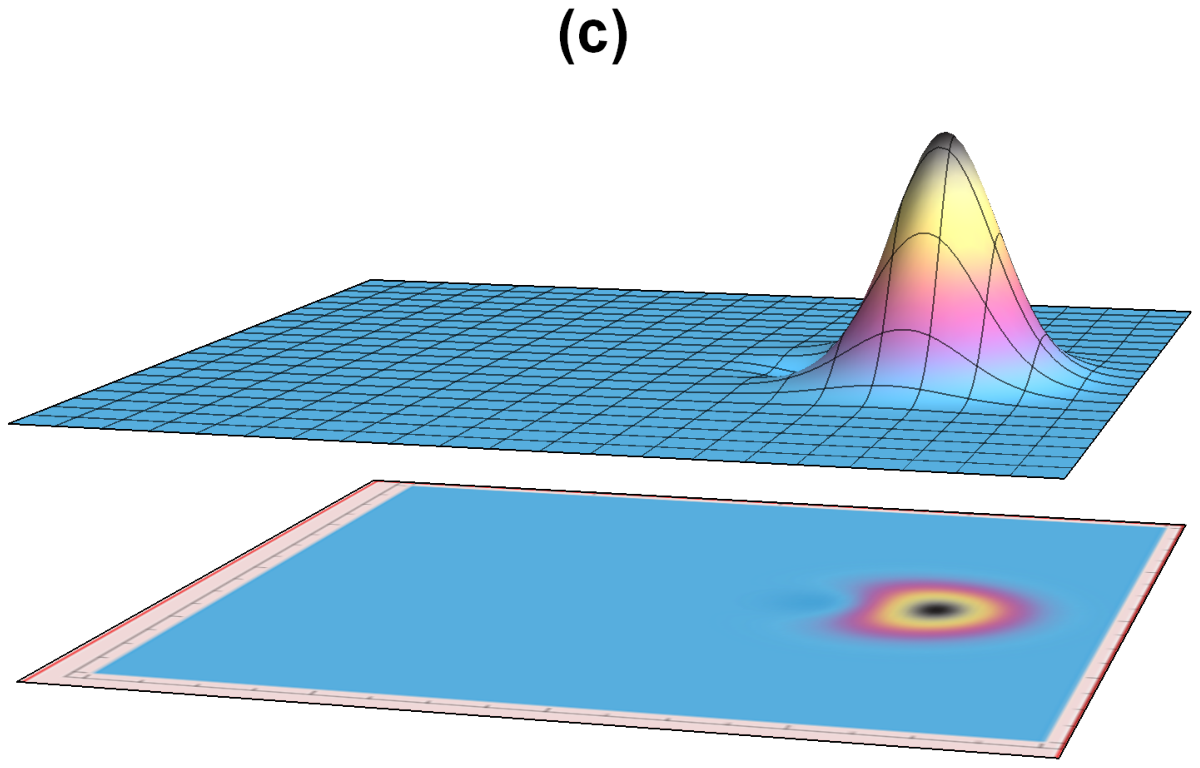}

\caption{\label{fig:4-1}(Color online) The Wigner function of output signal
state $\vert\Theta\rangle$. (a) $\vert\beta\vert=0$, (b) $\vert\beta\vert=1$,
(c) $\vert\beta\vert=2$. Other parameters are the same as Fig. \ref{fig:2}.}
\end{figure}

We can see that this Wigner function consists three parts. The first
and second terms corresponded to the Wigner function of coherent state
$\vert\beta\rangle$ and SPAC state $\text{\ensuremath{\vert1,\beta\rangle}}$,
respectively, and third term caused by their superposition. In Fig.
\ref{fig:4-1}, we plot the Wigner function of the state $\vert\varTheta\rangle$
for different amplitude $\beta$. From the Fig. \ref{fig:4-1}, we
can see that the negativity of $W\left(z\right)$ vanished gradually
with increasing the amplitude $\beta$. We know that every wave function
has its phase space distribution which characterized by Wigner function,
and it is an unique. This presented phenomena in Fig. \ref{fig:4-1}
is exactly the phase space distribution of SPAC state $\vert1,\beta\rangle$
\citep{PhysRevA.43.492}. Thus, when $\kappa_{2}\gg\kappa_{1}$, the
$\vert\varTheta\rangle$ gives us the new type of nonclassical state,
i.e., $\vert1,\beta\rangle$. 

\subsection{Signal to noise ratio (SNR)}

As shown in our schematic Fig. \ref{Fig:1}, the new output state
$\vert\varTheta\rangle$ of the signal beam is generated after we
taking the postselection to the idler beam which accomplished by D1
and D2. If we didn't take the postselection, then final state of the
signal will gives by Eq. ( \ref{eq:9-1}) after taking a trace to
the idler beam with state $\vert\psi_{i}\rangle$. However, since
in the nonpostselection case will not occur weak value of operators
$A$ and $B$ which possess the signal amplification feature, the
postselected weak measurement may have advantages over nonpostselected
measeurement in signal amplification process. To show the usefulness
of new generated state $\vert\varTheta\rangle$, here we study the
ratio of SNRs between the postselected and nonpostselected weak measurements
\citep{Agarwal2013}

\begin{equation}
\chi=\frac{\mathcal{R}_{X}^{p}}{\mathcal{R}_{X}^{n}}.\label{eq:31}
\end{equation}
 Here, $\mathcal{R}_{X}^{p}$ represents the SNR of postselected weak
measurement defined as
\begin{equation}
\mathcal{R}_{q}^{p}=\frac{\sqrt{NP_{s}}\delta q}{\sqrt{\langle q^{2}\rangle_{f}-\langle q\rangle_{f}^{2}}},\label{eq:32}
\end{equation}
 with 
\begin{equation}
\delta q=\langle\varTheta\vert q\vert\varTheta\rangle-\langle\beta\vert q\vert\beta\rangle.\label{eq:33}
\end{equation}
 Here, $N$ is the total number of measurements, $P_{s}$ is probability
of finding the postselected state for a given preselected state and
for our scheme it equal to $P_{s}=\vert\alpha\epsilon\vert^{2}$,
and $NP_{s}$ is the number of times the system was found in a postselected
state $\vert\psi_{f}\rangle$. Here, $\langle q\rangle_{f}$ denotes
the expectation value of measuring observable which defined in Eq.
( \ref{eq:5}) under the final state of the pointer (signal beam)
$\vert\varTheta\rangle$. 

When dealing with nonpostselected measurement, there is no postselection
process after the interaction between the system and pointer. Thus,
the definition of SNR for nonpostselected weak measurement can be
given as 
\begin{equation}
\mathcal{R}_{X}^{n}=\frac{\sqrt{N}\delta q^{\prime}}{\sqrt{\langle q^{2}\rangle_{f^{\prime}}-\langle q\rangle_{f^{\prime}}^{2}}},
\end{equation}
with 
\begin{equation}
\delta q^{\prime}=\langle\Psi\vert q\vert\Psi\rangle-\langle\beta\vert q\vert\beta\rangle.
\end{equation}
Here, $\langle q\rangle_{f^{\prime}}$ denotes the expectation value
of measuring observable under the final state of the pointer without
postselection which can be derived in Eq. ( \ref{eq:9-1}). In order
to evaluate the ration $\chi$ of SNRs, we have to calculate the related
quantities and related expressions are given as :

(1) the expectation value of $\langle q\rangle_{f}$ is 

\begin{align}
\langle q\rangle_{f} & =\langle\Phi\vert q\vert\Phi\rangle=\mathcal{N}\vert^{2}\{\vert\kappa_{1}\vert^{2}h_{1}+\vert\kappa_{2}\vert^{2}h_{2}-2Re[\kappa_{1}\kappa_{2}^{\ast}h_{3}]\},
\end{align}
where \begin{subequations}
\begin{align}
h_{1} & =\langle\beta\vert q\vert\beta\rangle=\sqrt{2}Re[\beta],\\
h_{2} & =\langle\beta\vert aqa^{\dagger}\vert\beta\rangle=\sqrt{2}(2+\vert\beta\vert^{2})Re[\beta],\\
h_{3} & =\langle\beta\vert aq\vert\beta\rangle=\frac{1}{\sqrt{2}}\left(1+\vert\beta\vert^{2}+\beta^{2}\right),
\end{align}
 \end{subequations} 

\begin{figure}
\includegraphics[width=8cm]{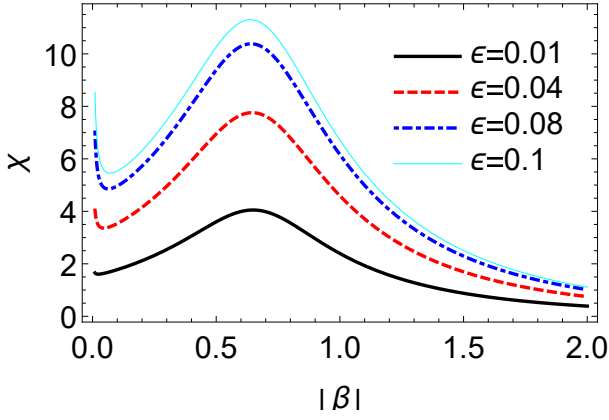}

\caption{\label{fig:3}(Color online) The ratio $\chi$ of SNRs between postselection
and nonpostselection weak measurement as a function of coherent state
parameter $\beta$ for different the slightly unbalanced parameter
$\epsilon$ of the beam splitter in our scheme. Other parameter are
the same as Fig. \ref{fig:2}.}
\end{figure}

(2) the expectation value of $\langle q^{2}\rangle_{f}$ is 

\begin{align}
\langle q^{2}\rangle_{f} & =\langle\Phi\vert q^{2}\vert\Phi\rangle=\vert\mathcal{N}\vert^{2}\{\vert\kappa_{1}\vert^{2}w_{1}+\vert\kappa_{2}\vert^{2}w_{2}-2Re[\kappa_{1}\kappa_{2}^{\ast}w_{2}]\},
\end{align}
where \begin{subequations}
\begin{align}
w_{1} & =\langle\beta\vert q^{2}\vert\beta\rangle=\frac{1}{2}\left(2Re[\beta^{2}]+2\vert\beta\vert^{2}+1\right),\\
w_{2} & =\langle\beta\vert aq^{2}a^{\dagger}\vert\beta\rangle=\frac{1}{2}\left(3+7\vert\beta\vert^{2}+2\vert\beta\vert^{4}+2(3+\text{\ensuremath{\vert\beta\vert^{2}}})Re[\beta^{2}]\right),\\
w_{3} & =\langle\beta\vert aq^{2}\vert\beta\rangle=\frac{1}{2}\left(3\beta+\beta^{3}+2\beta^{\ast}+\beta^{\ast}\vert\beta\vert^{2}+2\beta\vert\beta\vert^{2})\right).
\end{align}
 \end{subequations} The other quantities also can be obtained, and
here we didn't show all of them. The ratio of SNRs between postselected
and nonpostselected weak measurement is plotted as a function of coherent
state parameter $\beta$, and results are shown in Fig. \ref{fig:3}.
As we observed in Fig.\ref{fig:3}, the ratio $\chi$ is increased
and can be more larger then unity with increasing the unbalanced parameter
$\epsilon$ of the beam splitter for not very large $\vert\beta\vert$.
We have noticed that the magnitudes of weak values of $\hat{A}$ and
$\hat{B}$, Eq. ( \ref{eq:a}) and Eq. ( \ref{eq:b}) are inverse
to $\epsilon$. Thus, the small weak value is, the better postslected
SNR is achieved than nonpostselected one. In a word, it can draw a
conclusion that the postselected weak measurment can improve the SNR
rather than without post-selection one for smaller weak values.

\section{\label{sec:4}Generation of single- photon -added vacuum squeezed
state }

If assume the initial input state $\vert\phi\rangle$ of the signal
beam is prepared as SV state \citep{Walls} 
\begin{equation}
\vert\phi_{1}\rangle=S\left(\xi\right)\vert0\rangle,
\end{equation}
with $S\left(\xi\right)=\exp\left(\frac{1}{2}\xi a^{\dagger2}-\frac{1}{2}\xi^{\ast}a^{2}\right)$,
$\xi=\eta e^{i\varphi}$. Then, the output state of signal beam, Eq.
( \ref{eq:9}), becomes as 

\begin{align}
\vert\varOmega\rangle & =\chi\left(\vert\phi_{1}\rangle-\lambda_{1}a\vert\phi_{1}\rangle-\lambda_{2}a^{\dagger}\vert\phi_{1}\rangle\right).\label{eq:47}
\end{align}
Here $\lambda_{1}=\frac{g\alpha}{\sqrt{2}}$, $\lambda_{2}=\frac{g}{\sqrt{2}\alpha\epsilon},$
and
\begin{equation}
\chi^{-2}\!\!=\!1\!+\!\vert\lambda_{1}\vert^{2}\sinh^{2}\eta-\!Re[\lambda_{1}\lambda_{2}^{\ast}e^{i\theta}]\sinh(2\eta)+\!\vert\lambda_{2}\vert^{2}\cosh^{2}\eta\label{eq:50-1}
\end{equation}
is the normalization constant. In the below discussions, we can neglect
the term associated with the coefficient $\lambda_{1}$ since it is
too small compared to $\lambda_{2}$ for our allowed parameters. As
we can see, the state we prepared by optical modeling $\vert\Omega\rangle$
is the superposition of vacuum squeezed (VS) and single-photon-added
vacuum squeezed (SPAVS) states. These two states dominance depends
on the coefficients $\lambda_{1}$ and their amplitudes can be controlled
by beam spllitters and BBO crystal in our scheme (see Fig. \ref{Fig:1}).
In this section, by calculating the state distance, squeezing parameter
and Wigner function we proved that in allowed parameters region our
generated new state $\vert\Omega\rangle$ is very distinguished over
initial input state $\vert\phi_{1}\rangle$. 

\subsection{State Distance}

In order to investigate the similarities and differences of the generated
state $\vert\Omega\rangle$ between two states including SV state
and SPASV state, we evaluate the state distances between them.

1.The state distance between $\vert\Omega\rangle$ and squeezed vacuum
(SV) state$\vert\phi_{1}\rangle$ is given as 

\begin{align}
F_{1} & =\vert\langle\xi\vert\Psi\rangle\vert^{2}=\vert\chi\vert^{2}.\label{eq:49}
\end{align}

2. The state distance between $\vert\Omega\rangle$ and PASV state
$\vert\phi_{2}\rangle=\frac{a^{\dagger}\vert\xi\rangle}{\cosh\eta}$
is given as 

\begin{align}
F_{2} & =\vert\langle\phi_{2}\vert\varOmega\rangle\vert^{2}=\vert\frac{\chi}{\cosh\eta}\vert^{2}\vert\frac{1}{2}e^{i\theta}\lambda_{1}\sinh2\eta-\lambda_{2}\cosh^{2}\eta\vert^{2}.\label{eq:51}
\end{align}

\begin{figure}
\includegraphics[width=8cm]{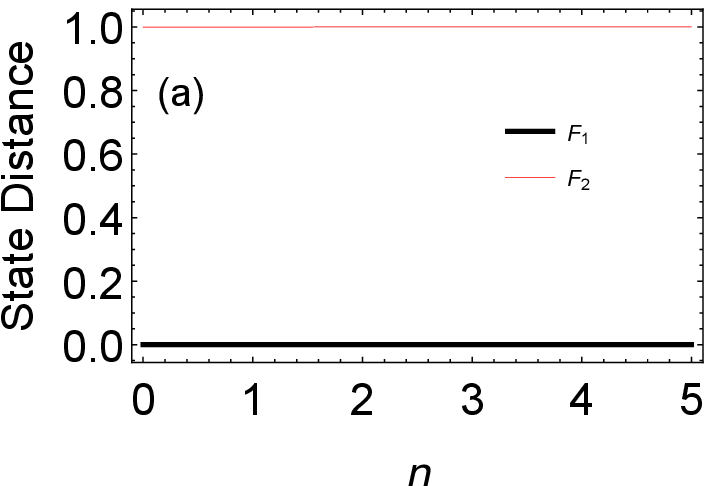}

\includegraphics[width=8cm]{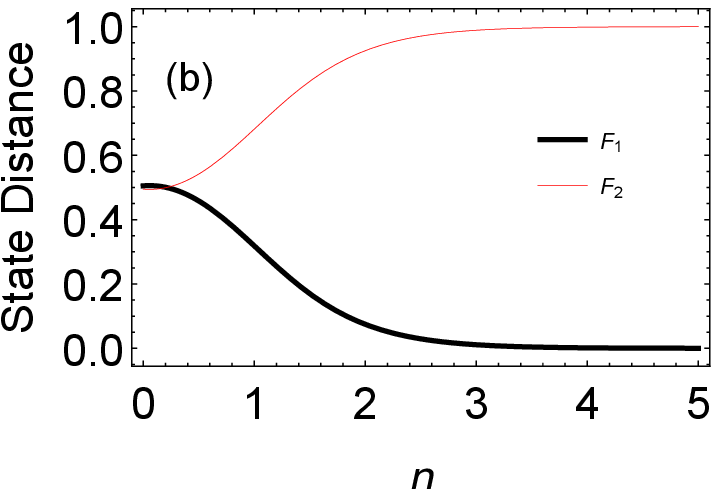}

\caption{\label{fig:6}(Color online) The state distance of $\vert\Omega\rangle\text{ between SVS and PASVS}$
as a function squeezed state parameter $\eta$ (a) for $\alpha=0.01$,
(b) for $\alpha=0.75$. Other parameters are the same as Fig. \ref{fig:2}.}
\end{figure}

In Fig. \ref{fig:6}, we plot separately the state distances between
$\vert\Omega\rangle$ and two states vs the squeezing parameter $\eta$.
As indictated in Fig. \ref{fig:6} (a), when the input idler coherent
state is too weak, $\alpha=0.01,$ the output signal state $\vert\Omega\rangle$
is very different with initial input state and the generated state
is totally same with the SPASV state. For $\alpha=0.75$, in very
weak squeezing parameter $\eta$, the output state $\vert\Omega\rangle$are
very similar to SV and SPASV states. But, as increasing the squeezing
parameter $\eta$, the similarities between $\vert\Omega\rangle$
and SPASV (SV) state is increased ( decreased) significantly (see
the Fig. \ref{fig:6}(b)). Although, the SPASV state only have one
photon difference between SV state, it have very different features
over SV state. Next we study the squeezing parameter and Wigner functions
of the new generated state $\vert\Omega\rangle$. 

\subsection{Squeezing parameter }

\begin{figure}
\includegraphics[width=8cm]{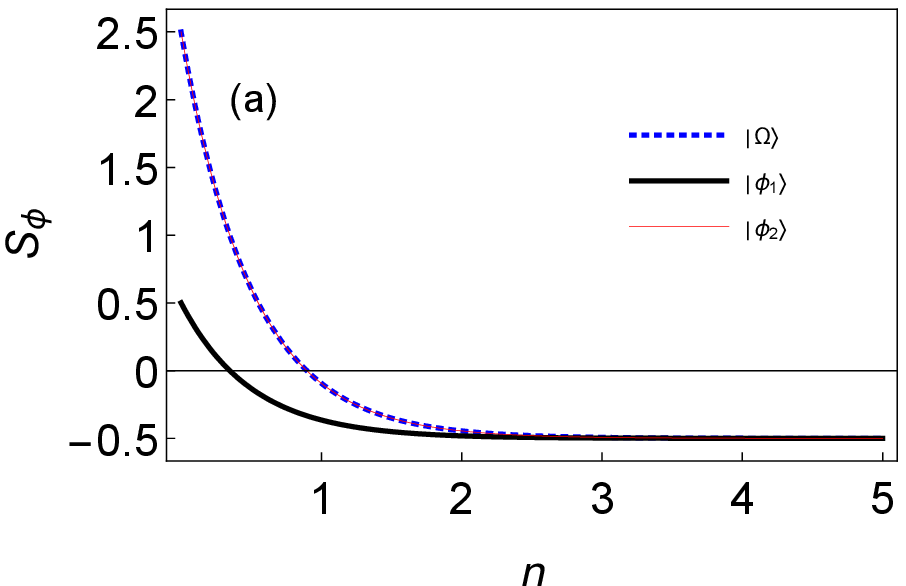}

\includegraphics[width=8cm]{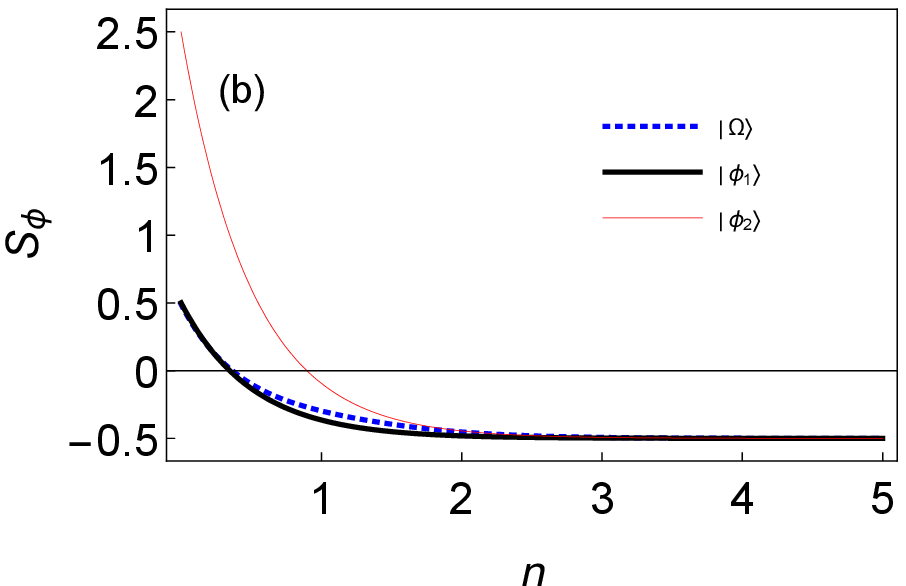}

\caption{\label{fig:8} (Color online) The squeezing parameter $S_{\phi}$
of $\vert\Omega\rangle$, SV state $\vert\phi_{1}\rangle$ and SPASV
state $\vert\phi_{2}\rangle$ vs squeezing parameter $\eta$ (a) for
$\alpha=0.01$, (b) for $\alpha=0.75$. Here we take $\phi=0$, and
other parameters are the same as Fig. \ref{fig:2}.}
\end{figure}

As we know, the SV state is an ideal state which possess very strong
squeezing effect. To investigate the squeezing effect of the field
quadrature of the generated state $\vert\Omega\rangle$, in this subsection
we study the squeezing parameter of $\vert\Omega\rangle$. The squeezing
parameter of radiation field is defined as 

\begin{equation}
S_{\phi}=(\triangle X_{\phi})^{2}-\frac{1}{2},\label{eq:50}
\end{equation}
 where 

\begin{equation}
\hat{X}_{\phi}=\frac{1}{\sqrt{2}}(ae^{-i\phi}+a^{\dagger}e^{i\phi}),\,\,\,\,\,\phi\in[0,2\pi],\label{eq:w}
\end{equation}
is the quadrature operator of the field, and $\triangle X_{\phi}=\sqrt{\langle\hat{X}_{\phi}^{2}\rangle-\langle\hat{X}_{\phi}\rangle^{2}}$
is the variance of variable $X_{\theta}$ . The minimum value of $S_{\phi}$
is $-0.5$ and if $-0.5\le S_{\phi}<0$ the field is called nonclassical.
We can calculate the squeezing parameter of SV state $\vert\phi_{1}\rangle$,
PASV state $\vert\phi_{2}\rangle$ and generated output state $\vert\Omega\rangle$
easily and their curves can be seen in Fig. \ref{fig:8}. We observe
from Fig. \ref{fig:8} (a) that when $\alpha=0.01$ the squeezing
parameter of the new generated output signal state $\vert\Omega\rangle$
exactly same with the squeezing parameter of SPASV state $\vert\phi_{2}\rangle$,
and it have very good squeezing as initial input state $\vert\phi_{1}\rangle$
when the squeezing parameter $\eta$ become larger. Furthermore, as
showed in Fig. \ref{fig:8}(b), if $\alpha=0.75$, then the squeezing
parameter of $\vert\Omega\rangle$ is same with the initial input
state $\vert\phi_{1}\rangle$ . Here, we have to mention that in our
scheme it is required that the measured system is initially prepared
in very weak coherent state. Thus, the $\alpha=0.75$ case is not
our main points. 

\subsection{Winger function of new generated state}

To further confirm similarities between SPASV state $\vert\phi_{2}\rangle$
and the new generated state $\vert\Omega\rangle$, in this subsection
we study the Wigner function of $\vert\Omega\rangle$. The Wigner
function for the state $\rho=\vert\Omega\rangle\langle\Omega\vert$
can be written as \citep{Int}
\begin{equation}
W(z)\equiv\frac{1}{\pi^{2}}\int_{-\infty}^{+\infty}\exp(\lambda^{\ast}z-\lambda z^{\ast})C_{W}(\lambda)d^{2}\lambda.\label{eq:35-1-1}
\end{equation}
Here, $C_{W}(\lambda)$ is the characteristic function, and is defined
as 
\begin{equation}
C_{W}(\lambda)=Tr\left[\rho e^{\lambda a^{\dagger}-\lambda^{\ast}a}\right],\label{eq:34-1}
\end{equation}
and $z=x+ip$ is represent the normalized dimensionless position and
momentum observables of the beam in phase space. After some math,
we can calculate the explicit expression of Wigner function of the
new generated state $\vert\Omega\rangle$, and it reads as 
\begin{align}
W(z) & =\vert\chi\vert^{2}w_{1}(z)+\vert\lambda_{1}\vert^{2}\vert\chi\vert^{2}w_{2}(z)+\vert\lambda_{2}\vert^{2}\vert\chi\vert^{2}w_{3}(z)\nonumber \\
 & -2\vert\chi\vert^{2}Im[\lambda_{1}e^{i\varphi}]w_{4}(z)-2Im[\lambda_{2}]\vert\chi\vert^{2}w_{5}(z)\nonumber \\
 & -2\vert\chi\vert^{2}Re[\lambda_{1}^{\ast}\lambda_{2}e^{-i\varphi}]w_{6}(z),\label{eq:53}
\end{align}
with \begin{subequations}
\begin{align}
w_{1}(z) & =\frac{2}{\pi}\exp\left[-2\vert\tilde{z}\vert^{2}\right],\\
w_{2}(z) & =\frac{2}{\pi}\sinh^{2}\eta\exp^{-2\vert\tilde{z}\vert^{2}}\left(4\vert\tilde{z}\vert^{2}-1\right),\\
w_{3}(z) & =\frac{2}{\pi}\cosh^{2}\eta\exp^{-2\vert\tilde{z}\vert^{2}}\left(4\vert\tilde{z}\vert^{2}-1\right),\\
w_{4}(z) & =\frac{4}{\pi}\mu\sinh\eta e^{-\tau},\\
w_{5}(z) & =\frac{4}{\pi}\mu^{\ast}\cosh\eta e^{-\tau},\\
w_{6}(z) & =\frac{1}{\pi}\sinh2\eta\exp^{-2\vert\tilde{z}\vert^{2}}\left(4\vert\tilde{z}\vert^{2}-1\right).
\end{align}
 \end{subequations}Here, $\tilde{z}=z\cosh\eta-z^{\ast}e^{i\theta}\sinh\eta$,
$\tau=2\Re^{2}[z](\cosh\eta-\sinh\eta)^{2}-2\Im^{2}[z](\cosh\eta+\sinh\eta)^{2}$
and $\mu=\Re[z](\sinh\eta-\cosh\eta)+i\Im[z](\sinh\eta+\cosh\eta)$.
We can observe that this Wigner function is a real function and its
value is bounded $-\frac{2}{\pi}\leq W(\alpha)\leq\frac{2}{\pi}$
in whole phase space. In the derivation of the above Wigner function
we have used the identities \begin{subequations} 
\begin{align}
S(\xi)aS^{\dagger}(\xi) & =a\cosh(\eta)-a^{\dagger}e^{i\varphi}\sinh(\eta),\\
S(\xi)a^{\dagger}S^{\dagger}(\xi)= & a^{\dagger}\cosh(\eta)-ae^{-i\varphi}\sinh(\eta).
\end{align}
\end{subequations}

\begin{widetext} 

\begin{figure}
\includegraphics[width=6cm]{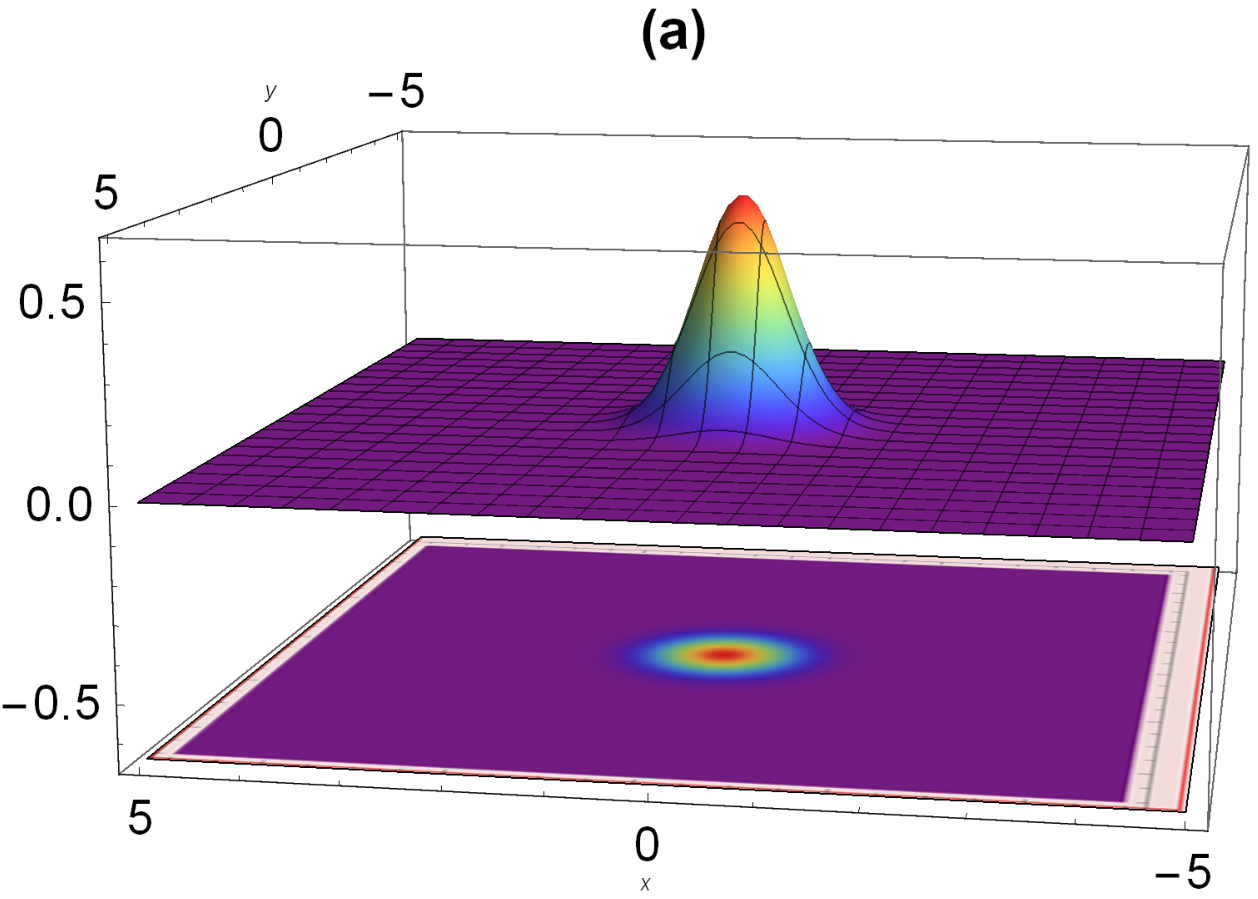}\includegraphics[width=6cm]{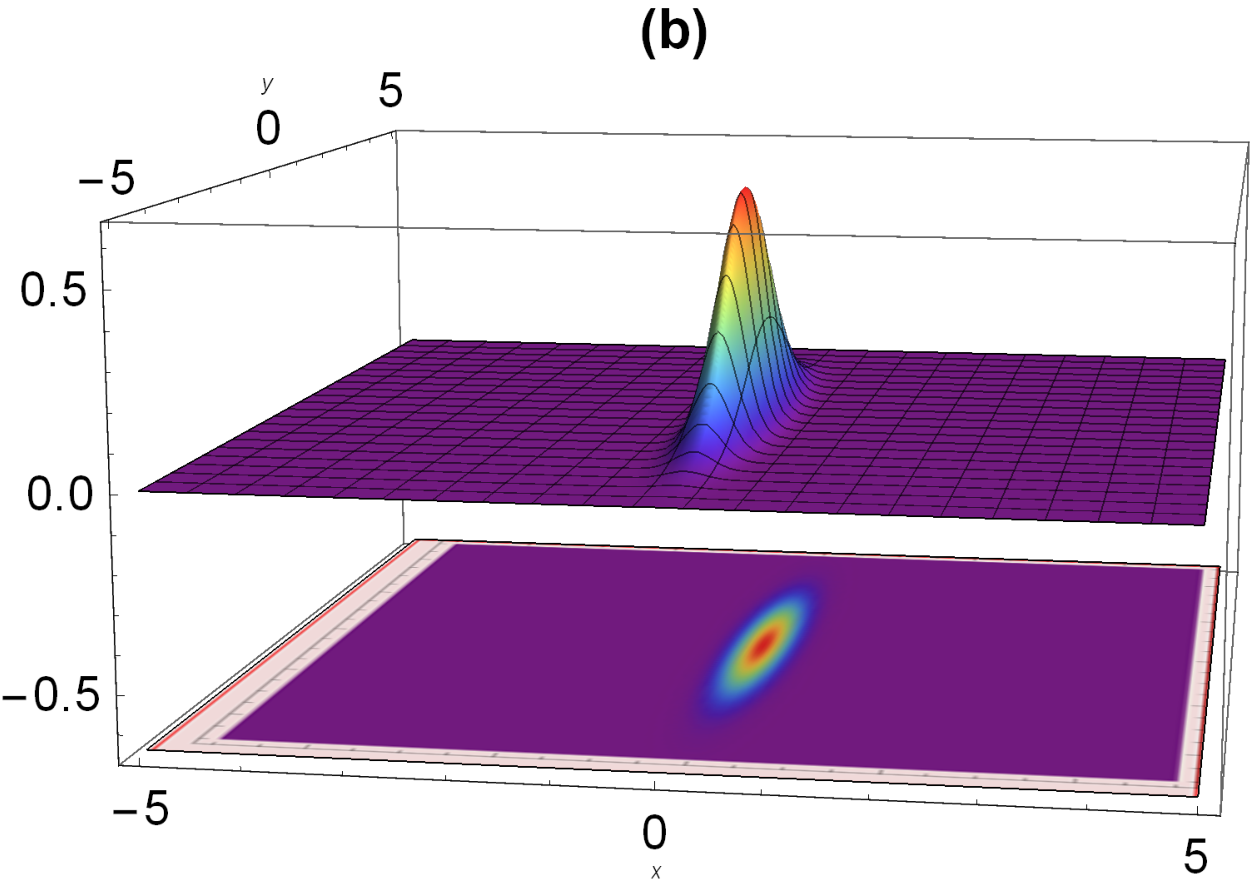}\includegraphics[width=6cm]{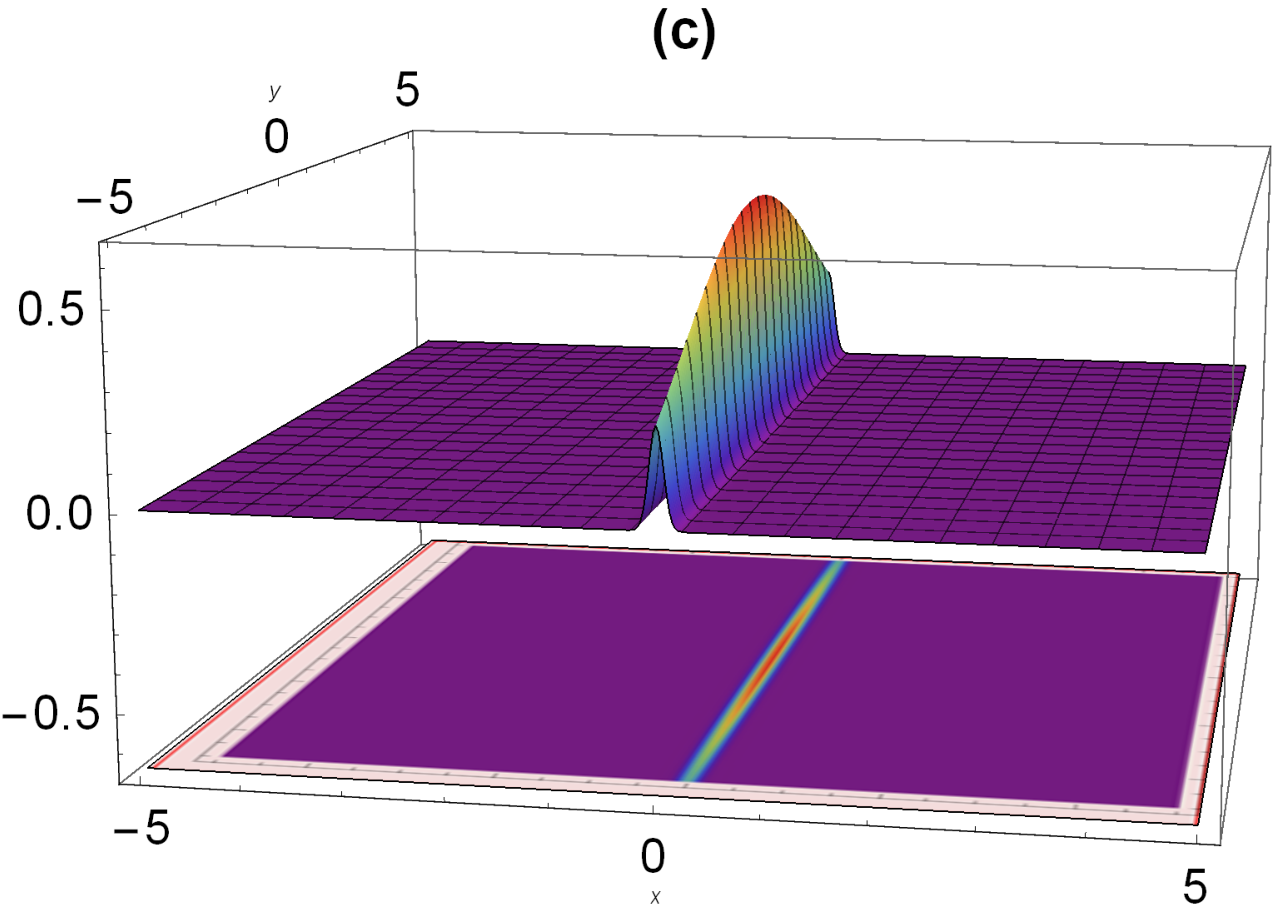}

\includegraphics[width=6cm]{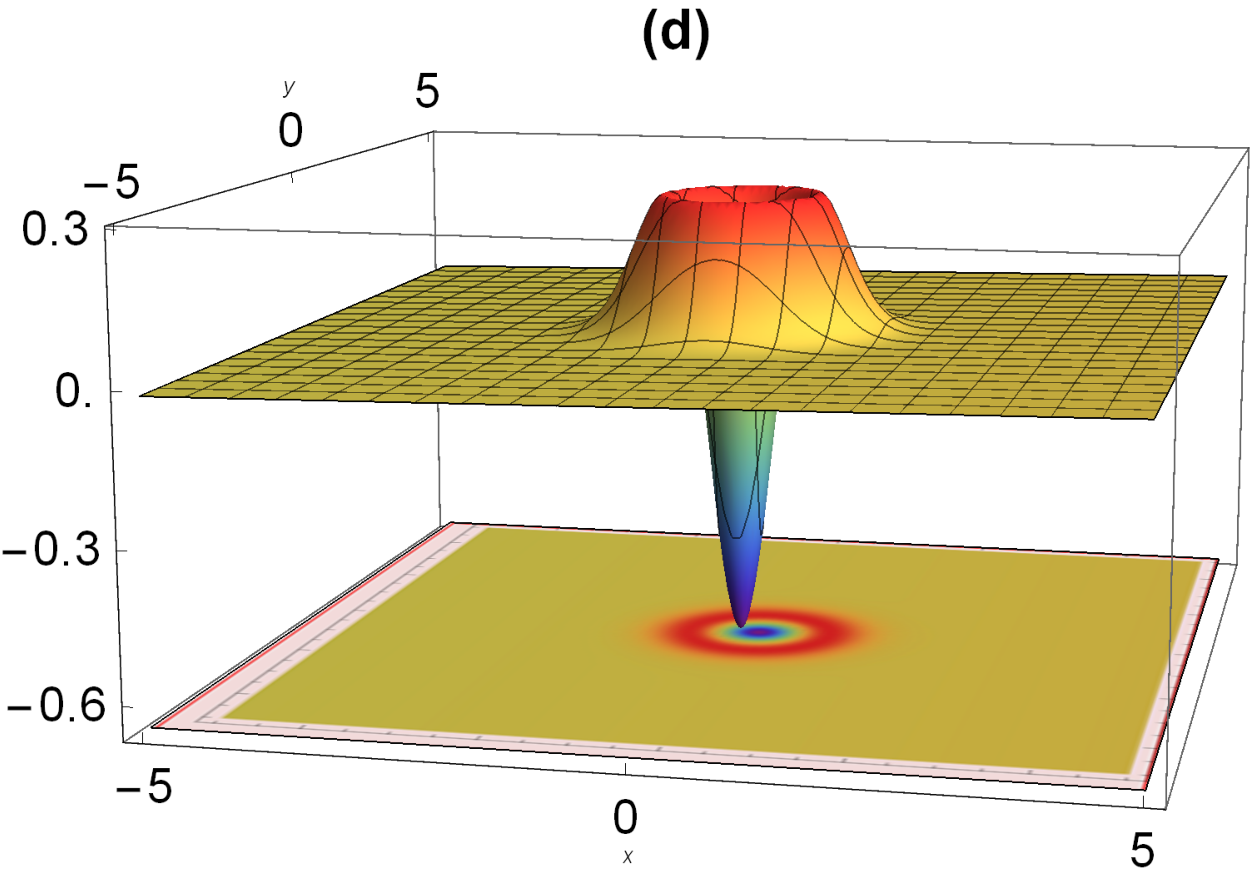}\includegraphics[width=6cm]{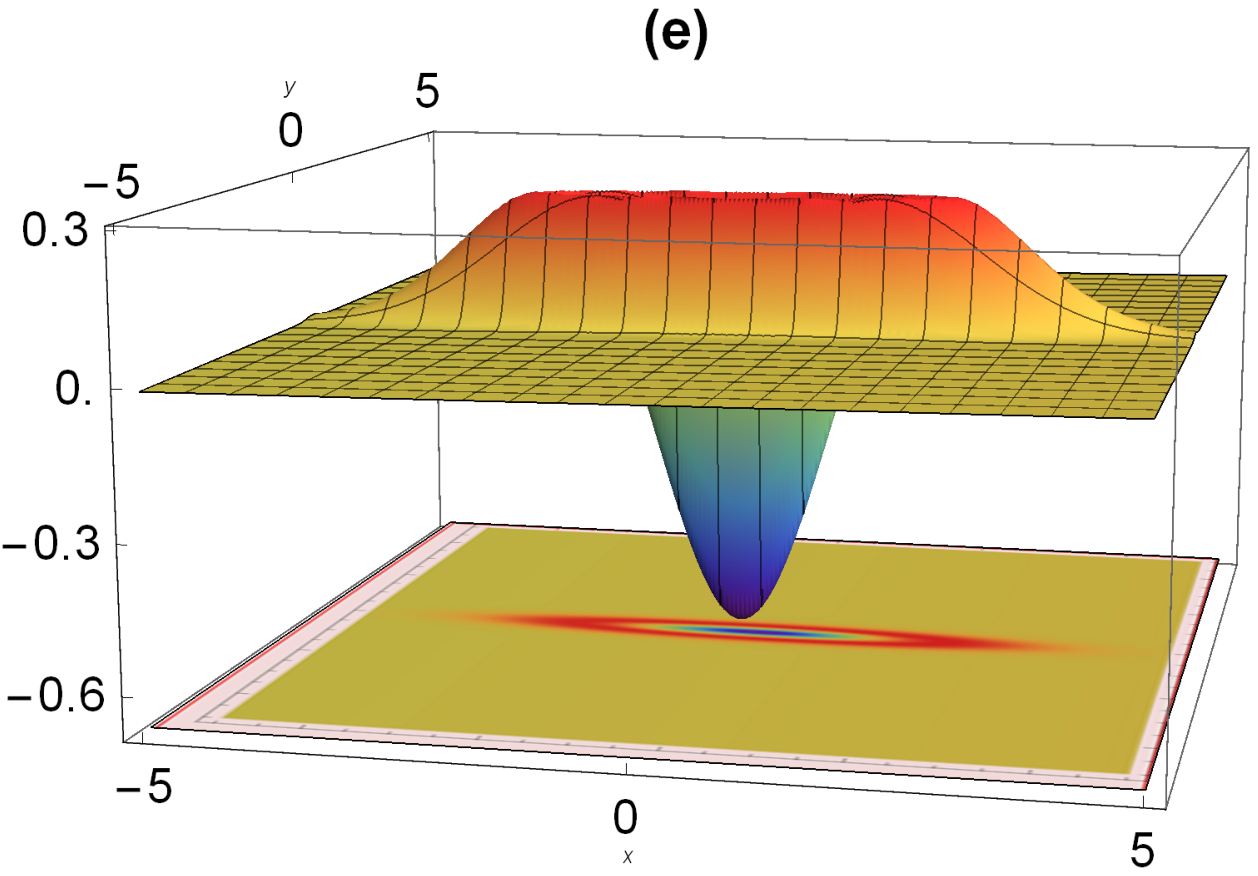}\includegraphics[width=6cm]{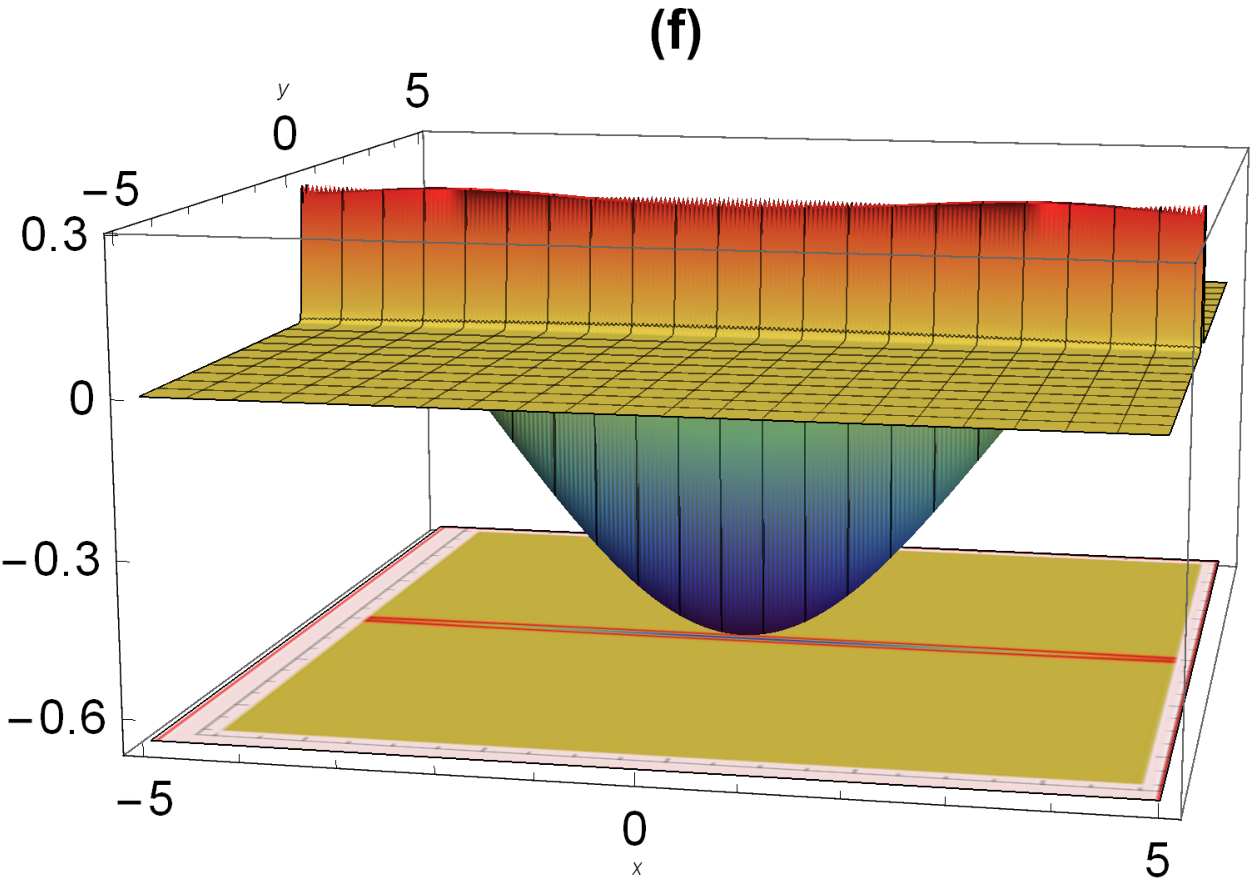}

\includegraphics[width=6cm]{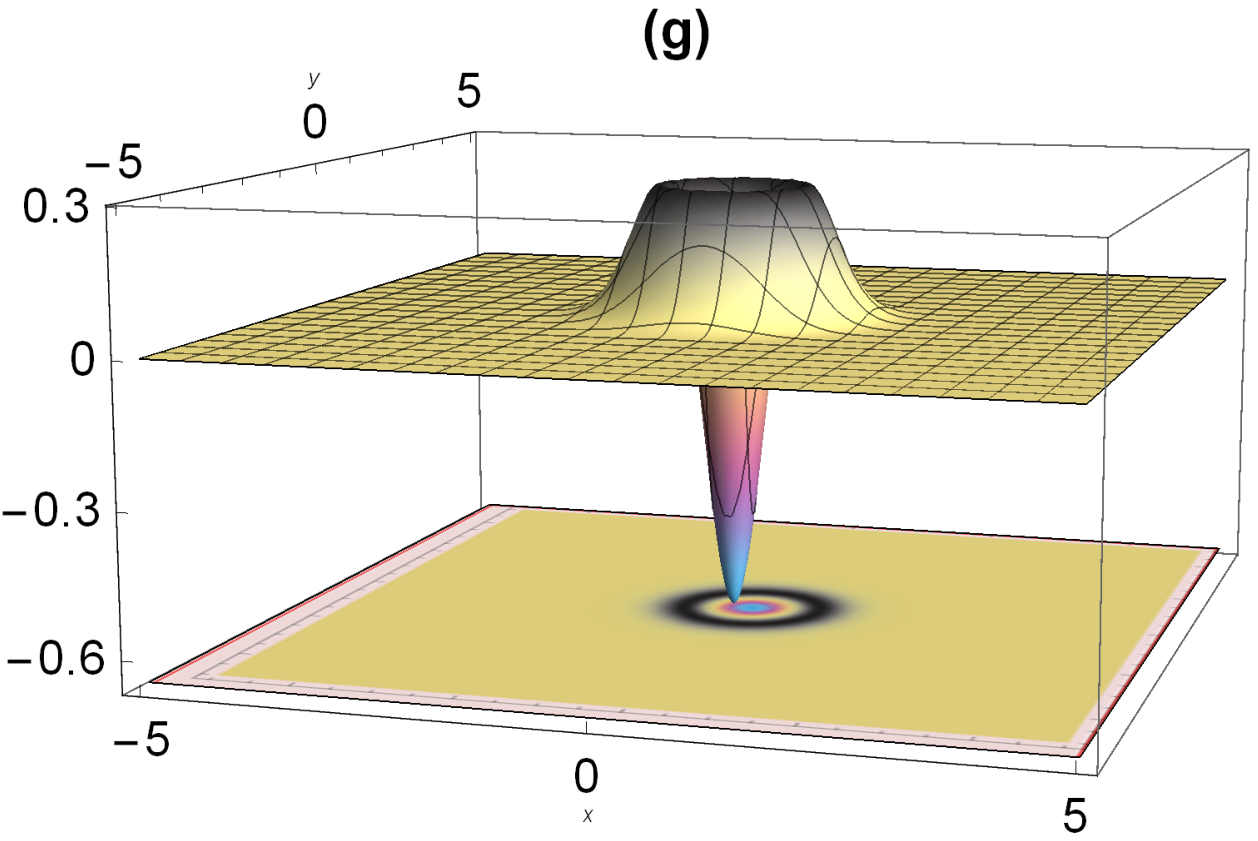}\includegraphics[width=6cm]{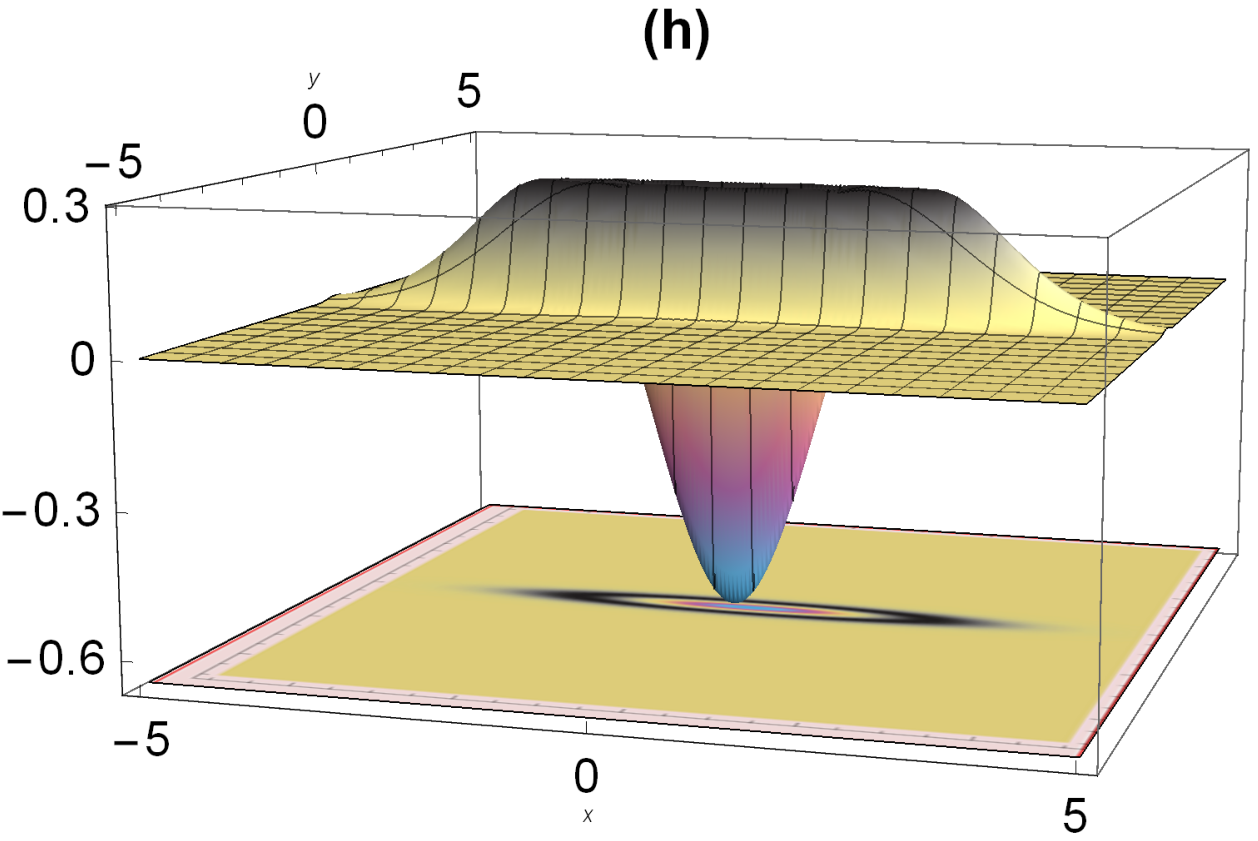}\includegraphics[width=6cm]{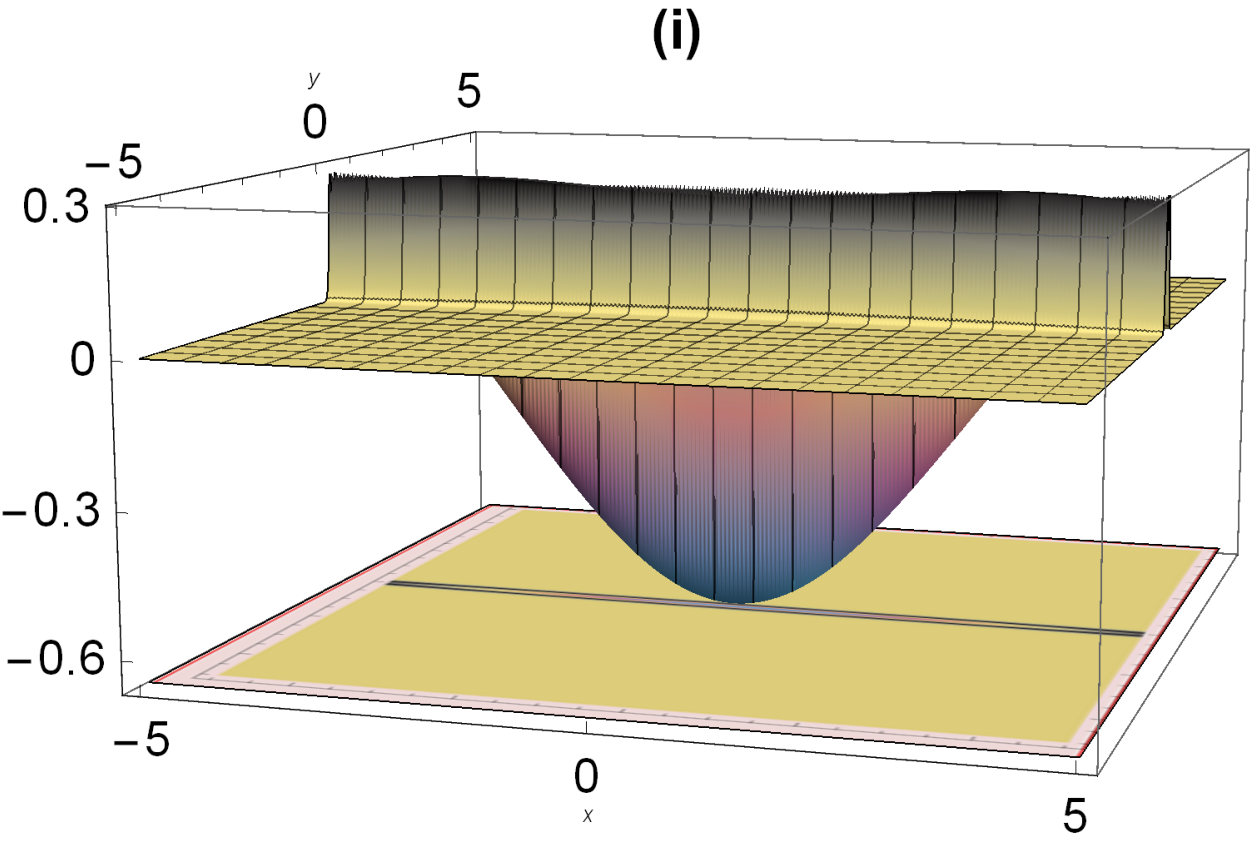}

\caption{\label{fig:9}(Color online) Comparing the Winger functions of SV
state $\vert\phi_{1}\rangle$, SPASV state $\vert\phi_{2}\rangle$
and new generated state $\vert\Omega\rangle$ in phase space. Each
column is defined for the different squeezing parameters $\eta$ (
set $\eta=0,1,2$), and are ordered accordingly from left to right.
Figures (a) to (c) correspond to the Winger functions of initial input
SV state $\vert\phi_{1}\rangle$, (d) to (f) correspond to the Winger
functions of new generated output signal state $\vert\Omega\rangle$,
and (g) to (i) correspond to the Winger functions of SPASV state $\vert\phi_{2}\rangle$,
respectively. Other parameters are the same as Fig. \ref{fig:2}.}
\end{figure}

\end{widetext}

The $w_{2}(z)$ in the Wigner function ( \ref{eq:53}) is the Winger
function of SV state $\vert\phi_{1}\rangle$. Although the SV state
$\vert\phi_{1}\rangle$ is a nonclassical state, its Winger function
is Gaussian and positive in phase space \citep{PhysRevA.75.032104}
. It is very clear in Eq. (\ref{eq:53} ) that it contains non-Gaussian
terms such as $w_{2}(z)$, $w_{3}(z)$ and $w_{6}(z)$. Thus the Wigner
function of our new generated signal state is non-Gaussian in the
phase space. We present the plots of the Winger functions of initial
input signal state $\vert\phi_{1}\rangle$ , new generated output
signal state $\vert\Omega\rangle$ and SPASV state $\vert\phi_{2}\rangle$
in phase space in Fig. \ref{fig:9} for different squeezing parameters
which we set as $\eta=0,1,2$. Figs. \ref{fig:9} (a)-(c) represents
Winger functions of SV state $\vert\phi_{1}\rangle$, and Fig. \ref{fig:9}
(d)-(f) represents Winger functions of new generated state $\vert\Omega\rangle$,
and Figs. \ref{fig:9} (g)-(i) represents Winger functions of SPASV
state $\vert\phi_{2}\rangle$, respectively. By comparing the curves
of those Winger functions, we observed that the generated state in
our scheme is a typical nonclassical state. It is very clear from
Fig. \ref{fig:9} (d)-(f) that as initial input state $\vert\phi_{1}\rangle$,
the new state $\vert\Omega\rangle$ has squeezing in one of the quadrature,
and there is also some negative regions of the Winger functions in
the phase space. These two features of the new state $\vert\Omega\rangle$
can show that its nonclassicality. Furthermore, it is proved that
our new generated state $\vert\Omega\rangle$ have exactly same phase
space distribution as SPASV state $\vert\phi_{2}\rangle$ (see second
and third rows of Fig. \ref{fig:9}). As indicated in Fig. \ref{fig:9}(d),
if the input state of the pointer is vacuum, then the output signal
state is prepared in single-phton Fock state.

\section{\label{sec:5}conclusion }

In summary, we have designed a fully laboratory feasible optical model
to successfully prepare nonclassical states such as single-photon
Fock state, SPAC state and SPASV state by using postselected weak
measurement in three wave mixing process. In our scheme the signal
and idler beams are taken as pointer and measured system, respectively,
and entanglement between them is realized by BBO crystal which can
take the role of weak measurement. In other words, in our study, a
nonlinear BBO crystal was chosen to introduce weak interaction between
three-wave mixing including pump, idle and signal light. By taking
the pre- and post-selections on measured system, the final pointer
state is prepared desired nonclassical state which depends on the
initial input signal state (initial pointer state). Further, we investigated
the properties including squeezing, second order correlation and Winger
functions of conditional output states.

We found that if the input signal (pointer) is vacuum state then the
output signal state is prepared in single-photon Fock state which
is typical quantum state exclusively used in many quantum information
processing. We also found that if the input signal state is coherent
(squeezed vacuum) state, then the output signal state prepared in
SPAC (SPAVS) state, respectively, and their purities can be controlled
by optical elements easily. Furthermore, we also found that the post-selective
measurement characterized by weak values and postselection have a
positive effect on the output SNR over non-postselection for coherent
state input case. 

Our scheme for the preparation of nonclassical states can be implemented
in optical Lab and we anticipate that this scheme could provide other
effective methods to the generation of other useful nonclassical state
such as Schrödinger kitten state \citep{3}. 
\begin{acknowledgments}
This work was supported by the National Natural Science Foundation
of China (Nos. 11865017, 11664041), the Natural Science Foundation
of Xinjiang Uyghur Autonomous Region (Grant No. 2020D01A72) and the
Introduction Program of High Level Talents of Xinjiang Ministry of
Science.
\end{acknowledgments}

\bibliographystyle{apsrev4-1}
\addcontentsline{toc}{section}{\refname}\bibliography{ref}

\end{document}